\documentclass[aps,prl,twocolumn,superscriptaddress,amssymb,floatfix]{revtex4-1}

\usepackage{dcolumn,bm,xstring,braket,amsmath,comment,nicefrac,placeins,color,graphicx,ragged2e}

\usepackage[unicode=true,pdfusetitle,bookmarks=true,bookmarksnumbered=true,bookmarksopen=false, breaklinks=false,pdfborder={0 0 0},pdfborderstyle={},backref=false,colorlinks=false, pdftitle={Spatial adiabatic passage of ultracold atoms in optical tweezers}]{hyperref}

\begin{document}

\title{Spatial adiabatic passage of ultracold atoms in optical tweezers}

\author{Yanay Florshaim, Elad Zohar, David Zeev Koplovich, Ilan Meltzer, Rafi Weill, Jonathan Nemirovsky, Amir Stern, and Yoav Sagi}

\email[Electronic address: ]{yoavsagi@technion.ac.il}

\affiliation{Physics Department and Solid State Institute, Technion --- Israel Institute of Technology, Haifa 32000, Israel}
\date{\today} 

\begin{abstract}
Spatial adiabatic passage (SAP) is a process that facilitates the transfer of a wave packet between two localized modes that are not directly coupled, but rather interact through an intermediate third mode. By employing a counter-intuitive adiabatic pulse sequence, this technique achieves minimal population in the intermediate state and high transfer efficiency. Here, we report the implementation of SAP for transferring massive particles between three micro-optical traps. We begin by preparing ultracold fermionic atoms in low vibrational eigenstates of one trap and then manipulate the distance between the three traps to execute the SAP protocol. We observe a smooth transfer of atoms between the two outer traps, accompanied by a low population in the central trap. We validate our findings and underscore the significance of the counter-intuitive sequence by reversing the order of the pulse sequence. Additionally, we investigate the influence of the tunneling rate and the time delay between the motion of the two external tweezers on the fidelity of the process. Our results open up new possibilities for advanced control and manipulation schemes in optical tweezer array platforms.
\end{abstract}

\maketitle

\emph{Introduction --} Modern quantum technologies require efficient and rapid control of the quantum state. Adiabatic following is one of the most effective methods for achieving this control by gradually connecting the initial and final states through a slow change of a specific parameter. In a two-level system coupled by an external field, continuously changing the drive frequency across the transition resonance leads to population inversion, a process known as adiabatic rapid passage (ARP) \cite{Allen1975}. A well-known extension of this concept to a three-level system in a lambda configuration is stimulated Raman adiabatic passage (STIRAP) \cite{Gaubatz1990}. STIRAP involves a counter-intuitive pulse sequence using two driving fields that connect the two low-energy states with the excited state. During STIRAP, the system is maintained in a dark state consisting only of the two lowest energy states. This is particularly advantageous in atomic systems with fast radiative decay of the excited state. Over the years, STIRAP has found wide-ranging applications in physics, chemistry, and engineering \cite{Vitanov2017}.

A particularly intriguing generalization of STIRAP involves three localized spatial modes that are coupled by tunneling \cite{Renzoni2001, Eckert2004, Greentree2004, MenchonEnrich2016}. This technique enables the transfer of a wave packet initially centered at the first mode to the third mode, with negligible wave packet population at the second mode throughout the process, despite a direct coupling of only the first and third modes to the second mode. Known as spatial adiabatic process (SAP), this process raises intriguing fundamental questions regarding the quantum continuity equation and the velocity of the transferred particle \cite{Benseny2012}. SAP can potentially be a valuable tool for efficiently relocating atoms within reconfigurable optical tweezer arrays. The possibility of observing coherent transport of a Bose-Einstein condensate was discussed in Ref.~\cite{Rab2008}. Experimentally, SAP has been demonstrated using light in photonic waveguides \cite{Longhi2006, Longhi2007, Valle2008, Longhi2009}. Transfer of ultracold atoms between different sub-lattices of a Lieb lattice was also recently reported \cite{Taie2020}. However, to date, the spatial adiabatic transfer of massive particles between localized potential wells has not been observed.

\begin{figure*}[t]
\begin{minipage}[t]{\linewidth}
 \includegraphics[width=\textwidth]{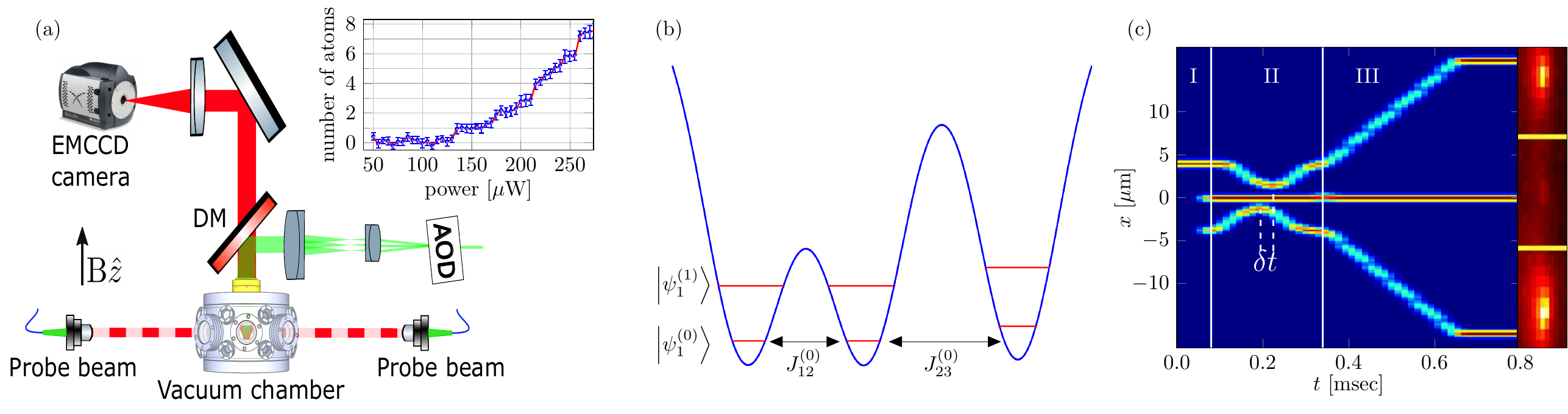}
\end{minipage}
\caption{\textbf{(a) Experimental setup.} The optical tweezers are created by diffracting a far-off-resonance laser beam from an acousto-optical deflector (AOD) at three different angles. These beams are then transmitted through an optical relay system and focused onto the plane of the atoms using an optical objective. The atoms plane is positioned at the center of the vacuum chamber, perpendicular to both the axis of the magnetic field and gravity. To collect the atomic fluorescence signal around 767 nm, generated by two counter-propagating on-resonance probe beams, the same objective collects the signal through a dichroic mirror (DM) and directs it to an electron multiplying charge-coupled device (EMCCD) camera. The inset shows the number of atoms versus the final tweezer power in the preparation process. Each point is an average of 35 measurements, and the red line is a guide to the eye.\textbf{(b) Diagram illustrating three identical tweezers potentials}, with two of them positioned closer to each other. Tunneling primarily occurs between eigenstates $\ket{\psi_i^{(n)}}$ with the same quantum number $n$. \textbf{(c) SAP sequence.} The spectrogram of the RF signal applied to the AOD, converted to the position in the atom's plane, is depicted. The sequence commences with the loading and preparation of tweezer 1 (I). Subsequently, tweezers 2 and 3 are turned on, and the counter-intuitive sequence is executed (II). At any given moment, the atomic occupancy in each tweezer can be measured by separating the two outermost traps to a distance of $40 \mu$m and conducting fluorescence imaging (III). An averaged image obtained from 500 runs is displayed on the right side.}
\label{fig:Sequence_of_preparation_and_detection}
\end{figure*}

In this paper, we present our observation of spatial adiabatic passage of a few fermionic atoms between three micro-optical traps (`optical tweezers'). Our experiment takes advantage of recent technological advancements in the dynamical control of optical tweezers \cite{Kaufman2021}. Initially, we arrange three optical tweezers in a line, with the atoms confined to only one of the external tweezers. By precisely manipulating the distance between the tweezers, we demonstrate that the atoms can be efficiently transferred to the other external tweezer, while the probability of finding them in the central tweezer remains low. We highlight the crucial role played by the counter-intuitive pulse sequence. Additionally, we investigate the tunneling rate between two adjacent tweezers and emphasize the significance of the relation between the tunneling rate and the tunneling coherence time as a key parameter in the SAP process.

\emph{SAP Model --} Each of the three optical tweezers is generated by a far-off-resonance laser beam, which creates a three-dimensional Gaussian potential through dipole interaction with the atoms \cite{Grimm2000}. In our experimental setup, the three beams propagate nearly parallel to each other, resulting in tunneling occurring primarily in the radial direction of the tweezer beams (see Fig. \ref{fig:Sequence_of_preparation_and_detection}a). Therefore, we can effectively treat the system as one-dimensional. During the preparation stage of the experiment, the atoms occupy the lowest vibrational eigenstates of the tweezer potential. For simplicity, we use tweezers with the same beam parameters and depth in this study. Consequently, eigenstates with the same principal quantum number $n$ at adjacent tweezers are degenerate (see Fig. \ref{fig:Sequence_of_preparation_and_detection}b). As long as the motion of the tweezers is adiabatic, tunneling only couples these degenerate states. Thus, for each $n$, we have a set of three coupled states denoted by $\ket{\psi^{(n)}_i}$ with $i \in {1,2,3}$. The initial state is $\ket{\psi(t=0)}=\ket{\psi^{(n)}_1}$.

The system evolves according to the time-dependent Schr\"{o}dinger equation, with the Hamiltonian \cite{MenchonEnrich2016}:
\begin{equation}
\label{eq:Hamiltonian}
\mathcal{H}(t) = \frac{\hbar}{2}
    \begin{pmatrix}
        0 & J^{(n)}_{12}(t) & 0 \\
        J^{(n)}_{12}(t) & 0 & J^{(n)}_{23}(t) \\
        0 & J^{(n)}_{23}(t) & 0
    \end{pmatrix},
\end{equation}
where $J^{(n)}_{ij}$ represents the real tunneling rate between the $n$-th eigenstates in traps $i$ and $j$. The Hamiltonian in Eq. \ref{eq:Hamiltonian} intentionally excludes direct tunneling between traps $1$ and $3$, in accordance with the SAP protocol. One particular eigenvector of interest is given by
\begin{equation}\label{eq:dark_state}
    \ket{D^{(n)}}=\cos{\theta^{(n)}}\ket{\psi^{(n)}_1}-\sin{\theta^{(n)}}\ket{\psi^{(n)}_3} \ \
\end{equation}
with an eigenvalue of $\lambda_D = 0$. This state is referred to as ``dark'' in the context of the original STIRAP scheme, as it does not include the excited state $\ket{\psi^{(n)}_2}$. The mixing angle $\theta^{(n)}(t)$ is defined as ${\tan{[\theta^{(n)}(t)]}={J^{(n)}_{12}(t)}/{J^{(n)}_{23}(t)}}$. By adiabatically varying $\theta(t)$ from $0$ to $\nicefrac{\pi}{2}$, the state $\ket{D^{(n)}}$ evolves from $\ket{\psi^{(n)}_1}$ to $\ket{\psi^{(n)}_3}$, allowing for the transport of the atom from tweezer $1$ to $3$ without ever occupying the middle tweezer. However, it's important to note that in any finite duration process which deviates from the ideal adiabatic description, there will also be some finite small occupation of the middle trap.

Initially, the three traps are well-separated from each other, and the coupling coefficients $J^{(n)}_{12}$ and $J^{(n)}_{23}$ are both zero. The counter-intuitive pulse sequence begins by increasing $J^{(n)}_{23}$ to a certain value, followed by an increase of $J^{(n)}_{12}$ while simultaneously decreasing $J^{(n)}_{23}$ \cite{MenchonEnrich2016}. The specific functional form of the time-dependent coupling rates is not crucial, as long as the process is adiabatic. In our experiment, we control these rates by adjusting the distances between the tweezers (see Fig.~\ref{fig:Sequence_of_preparation_and_detection}c). Notably, although the variation of the coupling coefficient with distance depends on the vibrational state $n$, the process can be adiabatic for a wide range of $n$ states. Since each set of $n$ vibrational states is isolated, we can perform the SAP experiment simultaneously with multiple atoms, each occupying a different eigenstate.

\emph{Experimental system --} Our experiment is conducted using a newly developed apparatus, which we will briefly describe (see Fig. \ref{fig:Sequence_of_preparation_and_detection}a). The apparatus consists of a spherical octagon stainless steel chamber with two re-entrant flanges with a 64 mm  diameter viewport, located above and below. These viewports allow for the placement of an optical objective outside the vacuum chamber at a distance of approximately 21 mm from the atomic plane, enabling a numerical aperture of 0.75. $\mathrm{^{40}K}$ atoms are dispensed in a separate glass chamber \cite{DeMarco1999}, where they are collected and guided into the main chamber using a 2D magneto-optical trap (MOT) \cite{Dieckmann1998}. The cooling sequence in the main chamber continues by capturing the atoms with a 3D MOT, followed by gray molasses cooling \cite{Salomon2013} and degenerate Raman sideband cooling \cite{Zohar2022}. The final cooling stage reduces the temperature to around 1 $\mathrm{\mu K}$ and optically pumps the atoms into the two lowest Zeeman eigenstates $m_{F}=-7/2,-9/2$ of the $F=9/2$ hyperfine manifold, with relative populations of 20\% and 80\%, respectively. The atoms are then loaded into a far-off-resonance crossed dipole trap (CDT) operating at a wavelength of 1064 nm with an aspect ratio of 1:2. After a brief optical evaporation period of 0.6 seconds, a radio frequency (RF) pulse is applied in a magnetic field of 185 G to make the spin mixture balanced. At this stage, the CDT holds approximately 80,000 atoms per spin state at a temperature of approximately twice the Fermi temperature.

At this stage, a single optical tweezer is turned on, overlapping with the Fermi gas. It is loaded with approximately 500 atoms per spin state. The phase space density is significantly enhanced thanks to the ``dimple effect'' \cite{StamperKurn1998}. As a result of the Fermi-Dirac distribution, there is nearly a unity probability of occupying the lowest eigenstates \cite{Serwane2011}. The optical tweezers are generated using a $1064$ nm laser beam that passes through an acousto-optical deflector (AOD), diffracting the beam to multiple angles based on the spectral content of its RF drive (see Fig. \ref{fig:Sequence_of_preparation_and_detection}a). These diffracted beams are optically guided to the optical objective, which focuses them to a Gaussian waist of $w_{0}=1.15\mu$m. The position and intensity of each tweezer can be dynamically controlled by adjusting the RF drive of the AOD. Once the tweezer is loaded, the crossed dipole trap (CDT) is gradually turned off, and the magnetic field is adjusted to its final value where the two spin states are non-interacting (208.7G) \cite{Shkedrov2018}. 

To remove atoms in high energy states, the intensity of the tweezer is reduced while simultaneously applying a magnetic field gradient of 2.5 G/cm along the $\hat{z}$ axis, which aligns with the axial direction of the tweezer and the direction of gravity \cite{Serwane2011,Spar2022}. This procedure allows control over the average number of atoms in the tweezer. In Fig.~\ref{fig:Sequence_of_preparation_and_detection}a) we present an example of measurement of the number of atoms in the tweezer as a function of the final tweezer power. Since there is a difference in the magnetic moment between the two lowest spin states of $\mathrm{^{40}K}$, there are typically two more atoms in the $m_F=-7/2$ spin state. However, the spin state has almost no relevance for the spatial adiabatic passage (SAP) process since, at the applied magnetic field, the spins are essentially non-interacting.

The preparation sequence is completed by gradually increasing the power of the tweezer. Subsequently, in a 1 ms interval, two additional tweezers are turned on and remain empty as the crossed dipole trap has already been turned off by this time. During the SAP sequence, a magnetic gradient is applied to cancel out the gravitational force. The gradient is set to the mean value between the two spin states, leaving approximately 5\% of the gravitational potential, which is negligible compared to the depth of the tweezer.

To measure the number of atoms in each tweezer, we reduce the magnetic field to 3G and apply two counter-propagating laser beams. These probe beams are resonant with the transition $\ket{\mathrm{F=9/2}} \rightarrow \ket{\mathrm{F'=11/2}}$. They have a linear polarization perpendicular to the direction of the magnetic field, enabling them to drive $\sigma_\pm$ transitions. To avoid interference and the formation of a standing wave pattern, the beams are turned on intermittently, each for a duration of 1 $\mu$s, with a total duration of 80 $\mu$s \cite{Bergschneider2018}. The photons scattered by the atoms are collected using the same high NA objective and directed onto an EMCCD camera (see Fig. \ref{fig:Sequence_of_preparation_and_detection}a).

The parameters of the tweezers are calibrated using two methods. Firstly, a direct imaging of the beam is performed to determine its size. Additionally, the harmonic trapping frequency is measured by modulating a piezoelectric actuator in one of the mirror mounts along the path of the tweezer beam, while observing the loss of atoms. Two resonant features can be observed, corresponding to the radial and axial directions. The measured harmonic trapping frequencies are found to be in agreement within 2\% of the values calculated based on the measured beam power and the imaged size of the tweezer. Furthermore, the number of atoms remaining in the tweezer is characterized as a function of the final potential depth \cite{Serwane2011}. To ensure that all three tweezers have an equal depth, a measurement is performed where the tweezers are positioned close enough to each other to allow for significant tunneling. The system is then allowed to reach equilibrium, and the depth of the tweezers is fine-tuned until the populations in each of the tweezers are equal.

\begin{figure}[!t]
\centering
\includegraphics[width=0.48\textwidth]{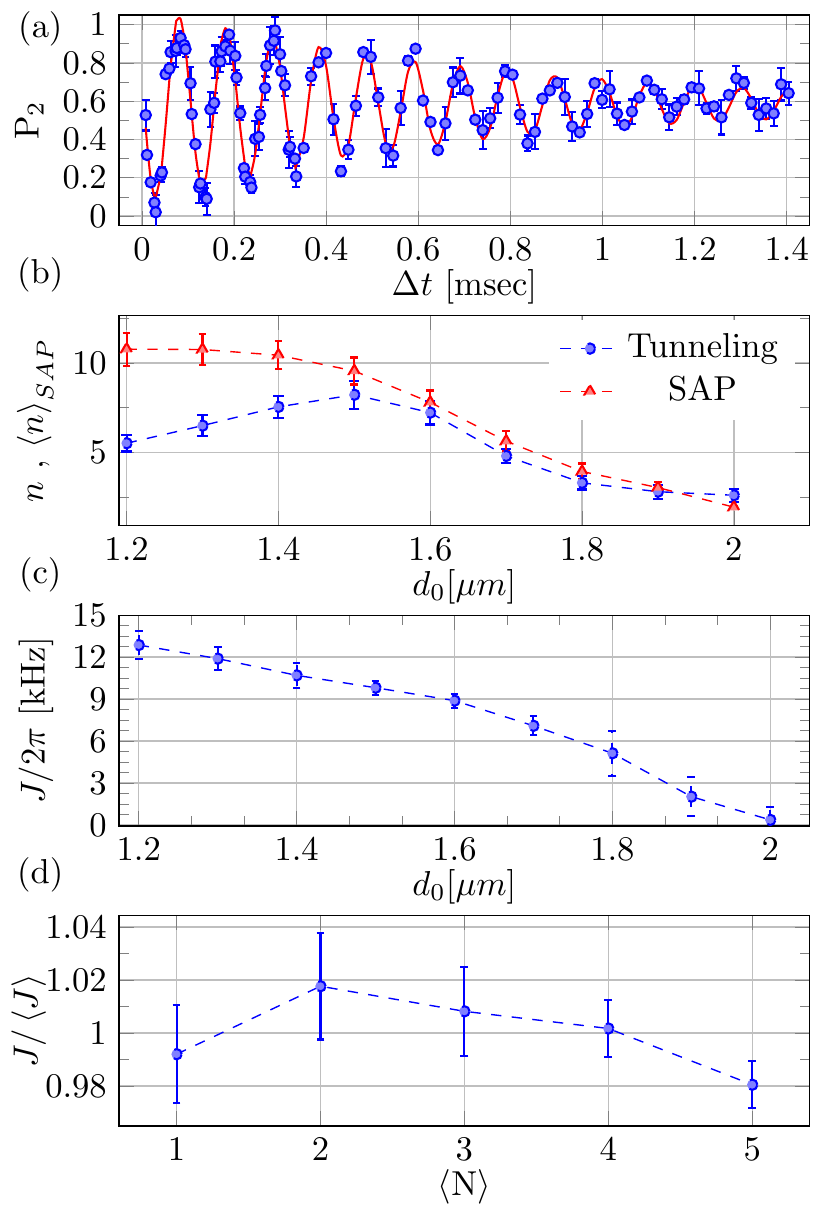}
\caption{\textbf{Tunneling measurements between two tweezers.} (a) The relative population in the initially empty tweezer versus the waiting time, $\Delta t$, is shown. The two tweezers are separated by a distance of $d_{0}=1.5\,\mu$m, and their depth is $U=k_B\times 95\,\mu$K, resulting in a calculated radial trapping frequency of $\omega_r= 2\pi \times 38.9\,\text{kHz}$. The data is fitted with a decaying sine function, $P_{2}(t)=c_1 e^{-t/\tau}\sin(Jt+\phi_0)+c_2$ (solid red line), from which the tunneling frequency $J$ and coherence time $\tau$ are extracted. (b) The number of coherent oscillations $n=J\tau/2\pi$ is plotted against the distance between the centers of the tweezers, $d_0$ (blue circles). Additionally, we plot the time average of $n$ during the SAP sequence, denoted as $\left \langle n \right \rangle _{SAP}=T^{-1}\int^{T}_{0} J(t)\tau(t)dt$ (red triangles). (c) The tunneling rate is shown as a function of $d_0$. (d) The relative variation of the tunneling frequency is plotted against the initial average number of atoms in the occupied tweezer, $\langle N \rangle$.}
\label{fig:tunneling}
\end{figure}

\emph{Tunneling measurements --}  We begin by investigating the tunneling behavior between adjacent tweezers, which is a key aspect of the SAP process. To conduct this measurement, we first prepare one tweezer with a small number of atoms, as described earlier. Subsequently, we gradually turn on a second empty tweezer at a distance of $d_0$ over a duration of 0.5 ms. After a certain waiting time $\Delta t$, we move the traps away from each other to a distance of 40 $\mu$m to terminate the tunneling, and then measure the relative population using fluorescence imaging (similar to Fig. \ref{fig:Sequence_of_preparation_and_detection}c).

An example of such a measurement, taken at a distance of $d_0=1.5\, \mu$m, is shown in Fig. \ref{fig:tunneling}a. Coherent oscillations between the tweezers, which decay slowly over time, are observed. The decay is likely attributed to noise in the tweezer position, fluctuations in its depth, and dephasing resulting from the involvement of multiple eigenstates of the two traps. Fig. \ref{fig:tunneling}b depicts the number of coherent oscillations as a function of the distance between the tweezers. We find an optimal distance of $d_0=1.5\, \mu$m, which corresponds to approximately 1.3 times the waist $w_0$. The maximum number of coherent oscillations we observe is similar to the findings reported for Rb atoms \cite{Kaufman2015}.

The tunneling frequency, $J/2\pi$, as a function of the distance between the tweezers is presented in Fig. \ref{fig:tunneling}c. As expected, the frequency decreases as the distance increases. However, we observe a relatively weak dependence on the distance that deviates from our expectation based on numerical simulations of the one-dimensional Schr{\"o}dinger equation for the ground state wave function. This discrepancy could be attributed to deviations in the shape of the tweezer from an ideal Gaussian, the influence of three-dimensional tunneling effects, and the population of higher vibrational states. Nonetheless, Fig. \ref{fig:tunneling}c provides valuable information regarding the relevant range of distances and the condition for adiabaticity in the SAP process. Furthermore, we conducted additional tunneling measurements with varying initial numbers of atoms, as depicted in Fig. \ref{fig:tunneling}d. The results demonstrate that the tunneling process is almost independent of the number of atoms, in this range of few atoms. This allows us to perform the measurement with few atoms simultaneously, thereby improving the signal-to-noise ratio. 

\emph{SAP experiments --} A schematic diagram illustrating the experimental SAP protocol is presented in Fig. \ref{fig:Sequence_of_preparation_and_detection}c. The protocol initiates by preparing tweezer no. 1 with an average of three atoms. At $t = 0$, tweezer no. 2 and tweezer no. 3 are simultaneously turned on within 0.5 ms at a distance where the tunneling is negligible. Since the CDT is already turned off at this stage, tweezers 2 and 3 are empty. All tweezers possess a depth of $95\,\mu$K and a Gaussian waist of $\omega_0 = 1.15\,\mu$m, identical to the tunneling measurements. In the counter-intuitive sequence, tweezer 3 initiates motion towards tweezer 2, and after a delay of $\delta t$, tweezer 1 follows the same trajectory towards tweezer 2. These trajectories modify $J_{12}(t)$ and $J_{23}(t)$, in accordance to the counter-intuitive sequence. The position of each of the external tweezers is given by a Gaussian profile 
\begin{equation}\label{Eq_movement_trajectory}
x_i(t)=\pm x_0 \pm (d_{min} - x_0)e^{-\frac{(t \mp \delta _t/2)^2}{2 \sigma ^2}}  \ \ ,
\end{equation}
where $i=1,3$ is the tweezer index, $x_0=4.5\,\mu$m is the initial position, $d_{min}$ is the minimal distance to the central tweezer, and $\sigma =0.194 T $ is the pulse width defined in terms of the total duration $T$. The upper (lower) sign pertains to tweezer $i=1$ ($i=3$). For clarity, Eq.(\ref{Eq_movement_trajectory}) is written such that $t=0$ is the middle of the process, but in the experiment we shift this time according to the sequence. At any given time throughout the sequence, the population in all three tweezers can be measured by swiftly moving ($\approx 15 \mu$sec) the two external tweezers back to their initial position where tunneling is essentially absent. This movement has been optimized to be faster than the tunneling dynamics while still maintaining a slow pace to prevent atom loss. Following this motion, the exact duration of the complete SAP sequence is awaited, after which the tweezers are moved again for imaging, as described earlier. The right panel of Fig.  \ref{fig:Sequence_of_preparation_and_detection}c presents an exemplary average fluorescence image of the three tweezers.

\begin{figure}
\centering
\includegraphics[width=0.48\textwidth]{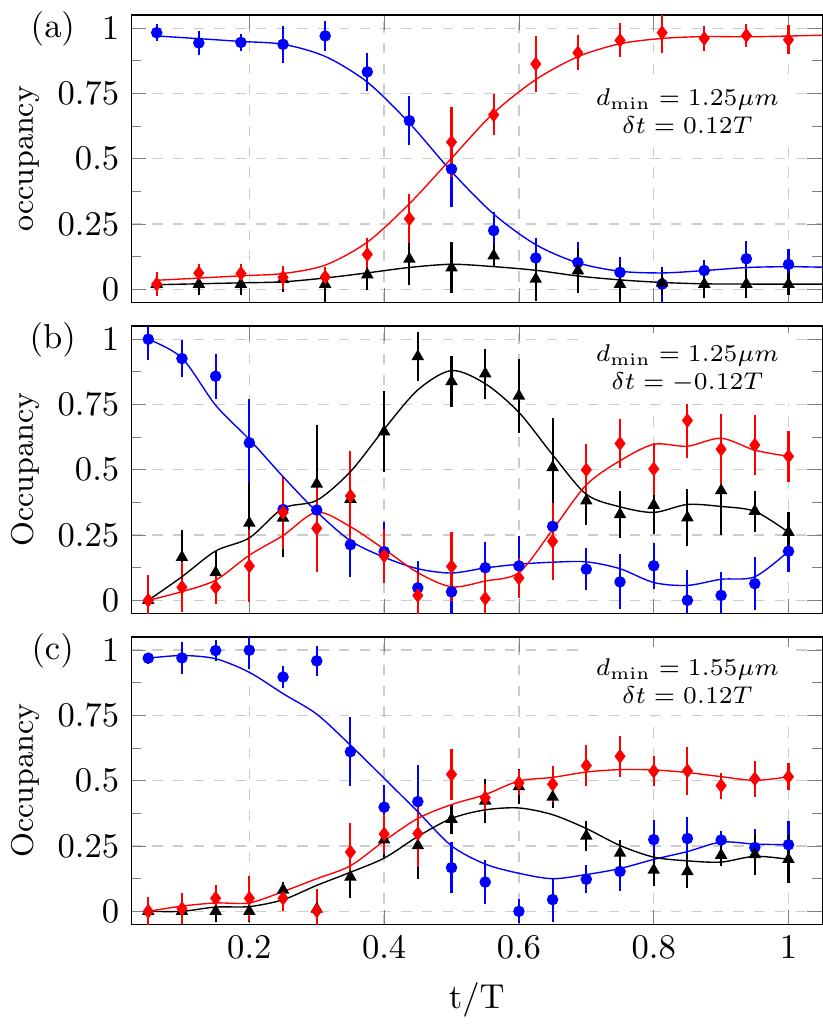}
\caption{\textbf{Measurements of SAP with three tweezers.} The relative population in tweezer 1 (blue circles), tweezer 2 (black triangles), and tweezer 3 (red diamonds) are plotted at different times during the SAP sequence. The solid lines serve as a visual aid. In panels (a) and (c) we plot a SAP process with a counter-intuitive profile ($\delta t >0$), while in (b) we use the intuitive sequence. The minimal distance between the tweezers is $d_{min} = 1.25\,\mu$m for (a) and (b) and $d_{min} = 1.55\,\mu $m in (c).}
\label{fig:SAP_vs_d_0}
\end{figure}

We present the main result of this paper, demonstration of a successful SAP scheme, in Fig.  \ref{fig:SAP_vs_d_0}a. The sequence has a total duration of $T = 0.25$ ms, a delay of $\delta t = 0.12T$, and a minimal distance of $d_{\text{min}} = 1.25\, \mu$m. The data clearly illustrates the efficient transfer of atoms from tweezer 1 to tweezer 3, while maintaining a minimal probability of occupation in tweezer 2. For comparison, we conducted an experiment with identical parameters but employing the intuitive pulse order, with $\delta t = -0.12T$. The contrasting result, displayed in Fig. \ref{fig:SAP_vs_d_0}b, is remarkably different from the counter-intuitive SAP depicted in Fig.  \ref{fig:SAP_vs_d_0}a. In the middle of the sequence, there is nearly complete atom transfer to tweezer 2. This outcome is expected in the intuitive sequence, which is close in spirit to a series of two $\pi$ pulses. Although there is a significant final population in tweezer 3, the transfer efficiency is lower compared to the counter-intuitive SAP, and notably, a finite population remains in the central tweezer.

An interesting question is how the SAP process depends on the minimum distance between the tweezers. In Fig.  \ref{fig:SAP_vs_d_0}c, we present the results of a similar experiment to that shown in Fig. \ref{fig:SAP_vs_d_0}a, but with a larger minimum distance of $d_{\text{min}} = 1.55,\mu$m. Surprisingly, in this case, the process is unsuccessful, leaving a substantial population in all three tweezers. This outcome may appear puzzling initially, considering that the largest number of coherent oscillations was achieved at this distance (Fig. \ref{fig:tunneling}b). However, it is crucial to note the distinction between the conditions in the tunneling experiments depicted in Fig. \ref{fig:tunneling} and those in the SAP process. During the SAP process, the distance between the tweezers undergoes changes. Consequently, a more appropriate measure would be to average the coherent tunneling oscillations over the entire SAP process. This quantity, represented by the red triangles in Fig.  \ref{fig:tunneling}b, indeed increases as the distance decreases. It is important to emphasize that we limit our investigation to $d_{\text{min}}>w_0$ to ensure there is a barrier between adjacent tweezers.

\begin{figure}
\centering
\includegraphics[width=0.48\textwidth]{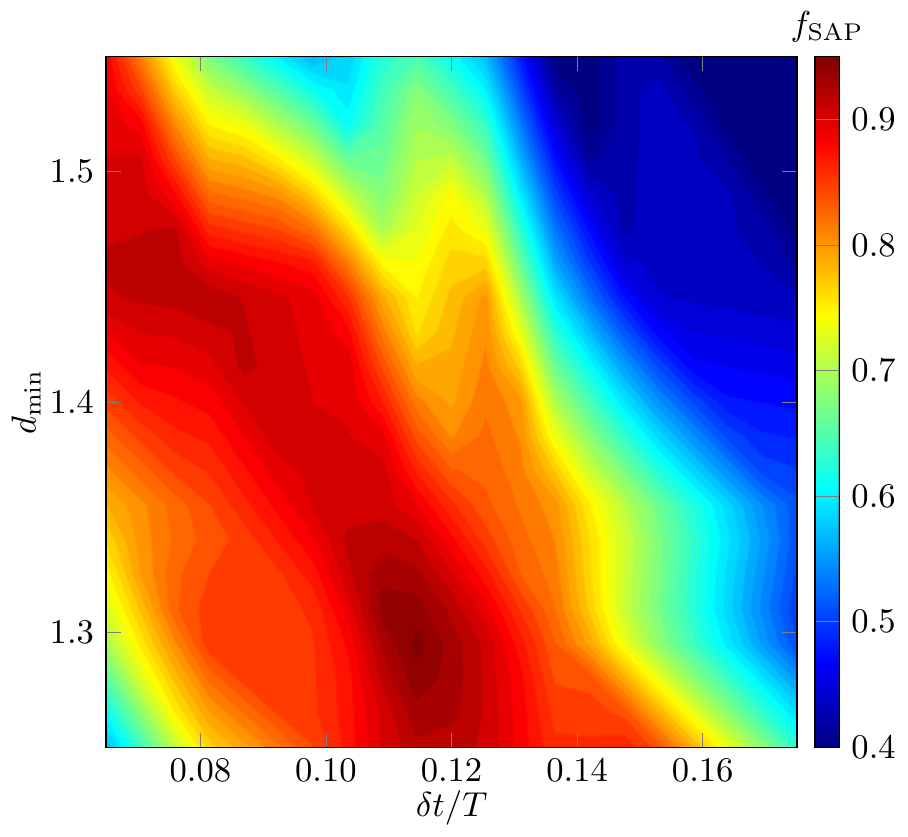}
\caption{\textbf{SAP fidelity}. Mapping of the $f_{SAP}$ versus the minimal distance, $d_{min}$, and pulses delay, $\delta t$.}
\label{fig:SAP_fidelity_d0_delay}
\end{figure}

In order to quantify the fidelity of the SAP process, we introduce the following function:
\begin{equation}
f_{\mathrm{SAP}}=P_{3}(T)\left (1-T^{-1}\int_{0}^{T}P_{2}(t)dt \right) \ \ ,
\label{eq1:fidelity_1}
\end{equation}
where $P_{i}(t)$ represents the probability of finding an atom in tweezer $i$ at time $t$. In an ideal counter-intuitive SAP process, $f_{\mathrm{SAP}}$ approaches unity. Fig. \ref{fig:SAP_fidelity_d0_delay}a illustrates the fidelity of the SAP protocol as a function of the minimal distance between the tweezers and the delay time. A high-fidelity operation is achieved over a wide range of parameters, as expected in an adiabatic process. The optimal time delay is approximately $0.12T$, which closely aligns with the theoretical value of $0.157T$ \cite{MenchonEnrich2016}.

\emph{Discussion --} In this work, we have successfully demonstrated the spatial adiabatic transfer of ultracold fermionic atoms between three optical tweezers. We have studied the tunneling dynamics between two tweezers. Our findings have revealed that the fidelity of the SAP process is ultimately limited by the coherence time of the tunneling dynamics. By achieving improved control over the potential landscape of the tweezers, it may be possible in the future to enhance the fidelity by increasing the number of tunneling oscillations. Additionally, the SAP protocol can be extended to encompass more than three tweezers \cite{Shore1991,Longhi2006a}. In the context of a chain of potential wells, the adiabatic passage exhibits characteristics reminiscent of Thouless pumping \cite{Lohse2015,Nakajima2016}, enabling robust transfer of topological edge states \cite{Mei2018,Longhi2019}. Therefore, SAP has the potential to serve as an essential building block in tweezer array quantum technology platforms.

\begin{acknowledgments}
We acknowledge helpful discussions with Yoav Lahini and Klaus Mølmer. This research was supported by the Israel Science Foundation (ISF), grant No. 3491/21, and by the Pazy Research Foundation. This research project was partially supported by the Helen Diller Quantum Center at the Technion.
\end{acknowledgments}

\bibliography{SAP_bib.bib}

\end{document}